\documentclass[aps,prd,psfig,amsmath,amsfonts,groupedaddress,eqsecnum]{revtex4}
\newcommand{\be}{\begin{equation}}
\newcommand{\ee}{\end{equation}}
\newcommand{\bea}{\begin{eqnarray}}
\newcommand{\eea}{\end{eqnarray}}
\newcommand{\bes}{\begin{subequations}}
\newcommand{\ees}{\end{subequations}}

\newcommand{\half}{\frac{1}{2}}

\begin{document}

%\input psfig.sty

% number in the upper righthand corner of the title page in preprint
%mode.
%\preprint{}

\title{Post-Newtonian gravitational radiation
and equations of motion via direct
integration of the relaxed Einstein equations. \\
V. Evidence for the strong equivalence principle to second post-Newtonian order}


\author{Thomas Mitchell and Clifford M. Will}
\email[]{cmw@wuphys.wustl.edu}
\homepage[]{wugrav.wustl.edu/people/CMW}
%\thanks{}
%\affiliation{Groupe Gravitation Relativiste et Cosmologie (GR$\varepsilon$CO)\\
%Institut d'Astrophysique, 98 bis Boulevard Arago, 75014 Paris, France}
\affiliation{McDonnell Center for the Space Sciences, Department of
Physics, \\Washington University, St. Louis, Missouri 63130
%\footnote{Permanent address}
}


\date{\today}

\begin{abstract}
Using post-Newtonian equations of motion for fluid bodies valid to the
second post-Newtonian order,
we derive the
equations of motion for binary systems with finite-sized, non-spinning but
arbitrarily shaped
bodies.  In
particular we study the contributions of the internal
structure of the bodies (such as self-gravity) 
that would diverge if the size of the bodies were to shrink to
zero.  Using a set of virial relations accurate to the first post-Newtonian
order that reflect the stationarity of each body, 
and redefining the masses 
to include 1PN and 2PN self-gravity terms, we demonstrate the complete
cancellation of a class of potentially
divergent, structure-dependent terms that
scale as $s^{-1}$ and $s^{-5/2}$, where $s$ is the characteristic size of
the bodies.  This is further evidence of the Strong
Equivalence Principle, and supports the use of post-Newtonian approximations
to derive equations of motion for strong-field bodies such as neutron stars
and black holes.  This extends earlier work done by Kopeikin.
\end{abstract}

\pacs{}
%\keywords{}

\maketitle

\section{Introduction and Summary }
\label{intro}

The principle of equivalence is the cornerstone of Einstein's general
theory of relativity.
Part of this principle, known as the weak
equivalence principle (WEP) states that test bodies fall in a
gravitational field with the same acceleration, irrespective of their
internal structure or composition.  By test body, one means a
body whose internal structure is governed by non-gravitational
interactions, and whose size is small compared to inhomogeneities in
external gravitational fields (suppression of tidal couplings).  
WEP, together with postulates of local Lorentz invariance
and local position invariance of non-gravitational physics in freely
falling frames, comprises the Einstein Equivalence Principle (EEP),
which is the foundation of metric gravity, or of the idea that gravity
is really geometry.  

There is a stronger version of WEP, which proposes that {\em all}
``test''
bodies should fall with the same acceleration, even bodies that are
self-gravitating, such as stars, planets, or black holes.  Here, by test
body, one means a body that is large and massive enough to have measurable
self-gravity, yet is small enough that tidal interactions can be
ignored (one generally ignores spin interactions as well).  This
version of WEP is a part of the Strong Equivalence Principle (SEP), which
includes a postulate of a kind of local Lorentz invariance and
position invariance of gravity itself.  While every metric theory of
gravity obeys EEP, almost no metric theory obeys SEP.  Scalar-tensor
theories of gravity, such as the Brans-Dicke theory and its
extensions, violate SEP.  Indeed, SEP is generically violated in
theories of gravity that introduce long-range fields, in addition to
the spacetime metric, that mediate how the metric is generated by
matter.  In order to preserve EEP, such fields do not couple directly
to matter.  Because general relativity contains one and only one
long-range gravitational field -- the metric itself -- it has no
mechanism for violating SEP.  Far from any gravitating system, the
metric can always be made suitably, if approximately, Minkowskian,
and so there is no obvious mechanism, other than tidal interactions, for the
external universe, or for any nearby ``spectator'' body to influence
the internal structure or dynamics of the system (for further details of
this argument, see \cite{nordwill} and Sec. 3.3 of \cite{tegp}). 
Thus gravity in GR
is independent of the velocity of the system relative to some external frame,
the effective constant
of gravity, $G$, is a true constant, and self-gravitating
non-spinning bodies move
as if they were test particles (for further discussion of SEP, see Section 3.3
of \cite{tegp}; for an
alternative discussion, see \cite{damour300}).  

But these are general, qualitative arguments.  
This paper
addresses the question: how {\em explicitly} does general relativity
manage to satisfy SEP for self-gravitating bodies, specificially to the
second order in a post-Newtonian expansion?

At the first post-Newtonian (1PN) order of approximation, that is, at order
$\epsilon \sim (v/c)^2 \sim Gm/rc^2$ 
beyond Newtonian gravity, GR has been shown to
obey SEP explicitly.  This is seen most graphically within the
parametrized post-Newtonian (PPN) framework, which characterizes the
post-Newtonian limit of a wide range of metric theories of gravity
using 10 arbitrary parameters (see Chapter 4 of \cite{tegp} for a review).  
Some of these parameters or combinations of
them measure
whether the theory has ``preferred-frame'' effects at post-Newtonian
order (violations of local Lorentz invariance), some measure whether
the locally measured gravitational constant can depend on the presence
of nearby matter (violations of local position invariance), and some
measure whether self-gravitating bodies violate WEP, a phenomenon known
as the Nordtvedt effect.  In GR all such offending parameters or
combinations of parameters vanish.  Furthermore, there is experimental
evidence that they vanish from a variety of tests of
post-Newtonian gravity, spanning 
lunar laser ranging, geophysical measurements and binary pulsar data (see
\cite{livrev} for a review).

But what about beyond post-Newtonian order?  Does the motion of a pair
of finite sized, gravitationally bound, non-spinning bodies depend on
their internal structure, apart from tidal interactions?  There is
existing theoretical evidence that they do not.  
Breuer and Rudolf~\cite{breuer} 
showed
that the relative ``Newtonian-like'' acceleration of two well separated 
bodies, momentarily at rest, was independent of their internal structure, to
second post-Newtonian
(2PN) order, or order $\epsilon^2$ beyond Newtonian gravity; 
put differently, they proved the absence of the Nordtvedt effect
to that order.
Kopeikin~\cite{kopeikin85} derived the equations of motion to 2PN order for
self-gravitating bodies with an equation of state $p(\rho)$, 
under the assumptions that they were spherically
symmetric in a suitable comoving frame, and had no internal fluid motions,
and also found that the SEP held (see also~\cite{GK86}).

This question is of more than academic interest.  The leading
candidate sources for gravitational radiation that may be detected in the
near future by
laser interferometers are binary systems containing neutron stars or
black holes.      
The inspiral part of their evolution can only be described accurately by the
post-Newtonian approximation, which, by its very nature, assumes that
gravitational fields are weak everywhere in spacetime.  No attempt has ever
been made to derive the motion to high orders in a PN expansion with either
strong-gravity neutron stars or black holes.  The closest one has come is to
treat the sources as distributions in spacetime (delta functions) and to
employ suitable regularization techniques to control the
infinities~\cite{blanchetfaye1,blanchetfaye2}, 
or to use a variant of the surface integral approach pioneered by Einstein,
Infeld and Hoffmann~\cite{futamase00}.
Therefore an explicit demonstration that the internal structure is
irrelevant for the binary motion, even if only at 2PN order, would be
valuable.

In this paper we derive the equations of motion of two arbitrarily
shaped, non-spinning, gravitationally bound bodies, 
through 2PN order.  Beginning with the 2PN hydrodynamic equations of motion
derived in Paper I of this series \cite{dire1}, 
we calculate the acceleration of the center of
``baryonic'' mass of a given body in the field of a companion body.
Expanding all variables about the centers of mass of each body we track
all
terms that scale with the sizes $s$ of the bodies as $s^{-1}$, and
$s^{-5/2}$; such terms represent contributions due to self-gravitational
binding energy, and a kind of gravitomagnetic internal energy,
respectively.  (In Paper II \cite{dire2} we ignored all finite-size
effects in deriving binary equations of motion.)  
We then employ virial relations that reflect the
equilibrium state of each body,  leading to a cancellation of many
terms.  Finally we renormalize the masses to include
both rest mass and gravitational binding energy, by defining the total
mass for body A to be
\be
M_A \equiv m_A + \frac{1}{2} {\hat \Omega}_A + O(m_A \epsilon^2) \,,
\ee
where $m_A = \int_A \rho^* d^3x$ is the conserved baryonic mass of body A
and
\be
{\hat \Omega}_A \equiv - \frac{1}{2} \int_A
\int_A \frac{{\hat \rho}^* {\hat \rho}^{*\prime}}
{|{\hat {\bf x}}- {\hat {\bf x}}^\prime|} d^3{\hat x} d^3{\hat
x}^\prime
\label{omegahat0}
\ee
is the gravitational self-energy {\em as measured in a local,
freely falling frame} ${\hat x}^\alpha$, momentarily comoving with
body
A.
The result is that
{\em all} self-gravity terms that scale as $s^{-1}$ or $s^{-5/2}$
cancel, leaving equations of motion at 2PN order for body 1 that
depend on the renormalized masses $M_A$, given by
\be
a_1^i = -\frac{M_2}{r^2} n^i + a_{1\,(1PN)}^i +a_{1\,(2PN)}^i \,,
\ee
where
$r= |{\bf x}_1 -{\bf x}_2|$, $n^i = (x_1-x_2)^i/r$, and where
\allowdisplaybreaks{
\bes
\bea
a_{1\,(1PN)}^i &=&
\frac{M_2}{r^2} n^i \left [
4\frac{M_2}{r} + 5\frac{M_1}{r} - v_1^2 +4 {\bf v}_1 \cdot {\bf v}_2
- 2 v_2^2 + \frac{3}{2}({\bf v}_2\cdot {\bf n})^2 \right ]
\nonumber
\\
&& + \frac{M_2}{r^2} (v_1 - v_2)^i
\left [ 4 ({\bf v}_1\cdot {\bf n}) -3 ({\bf v}_2\cdot {\bf n})\right ]
\,,
\label{1PNeomfinal}
\\
a_{1\,(2PN)}^i &=& \frac{M_2}{r^2}n^i\bigg[
\frac{M_2}{r}\bigg(
4v_2^2
-8{\bf v}_1\cdot {\bf v}_2
+2({\bf v}_1\cdot {\bf n})^2
-4({\bf v}_1\cdot {\bf n})({\bf v}_2\cdot {\bf n})
-6({\bf v}_2\cdot {\bf n})^2
\bigg)
\nonumber\\
&&
+\frac{M_1}{r}\bigg(
\frac{5}{4}v_2^2
-\frac{5}{2}{\bf v}_1\cdot {\bf v}_2
-\frac{15}{4}v_1^2
+\frac{39}{2}({\bf v}_1\cdot {\bf n})^2
-39({\bf v}_1\cdot {\bf n})({\bf v}_2\cdot {\bf n})
+\frac{17}{2}({\bf v}_2\cdot {\bf n})^2
\bigg)
\nonumber \\
&&
-\frac{57}{4}\frac{M_1^2}{r^2}
-\frac{69}{2}\frac{M_1M_2}{r^2}
-9\frac{M_2^2}{r^2}
-2v_2^4
+4v_2^2({\bf v}_1\cdot {\bf v}_2)
-2({\bf v}_1\cdot {\bf v}_2)^2
\nonumber\\
&&
+\frac{3}{2}v_1^2({\bf v}_2\cdot {\bf n})^2
-6({\bf v}_1\cdot {\bf v}_2)({\bf v}_2\cdot {\bf n})^2
+\frac{9}{2}v_2^2({\bf v}_2\cdot {\bf n})^2
-\frac{15}{8}({\bf v}_2\cdot {\bf n})^4
\bigg]
\nonumber \\
&&
+\frac{M_2}{r^2}(v_1^i - v_2^i)\bigg[
\frac{M_1}{4r}\bigg( {55}({\bf v}_2\cdot {\bf n})
-{63}({\bf v}_1\cdot {\bf n})\bigg)
-\frac{2M_2}{r}\bigg(({\bf v}_1\cdot {\bf n})
+({\bf v}_2\cdot {\bf n})\bigg)
\nonumber \\
&&
+v_1^2({\bf v}_2\cdot {\bf n})
+4v_2^2({\bf v}_1\cdot {\bf n})
-5v_2^2({\bf v}_2\cdot {\bf n})
-6({\bf v}_1\cdot {\bf n})({\bf v}_2\cdot {\bf n})^2
\nonumber \\
&&
-4({\bf v}_1\cdot {\bf n})({\bf v}_1\cdot {\bf v}_2)
+4({\bf v}_2\cdot {\bf n})({\bf v}_1\cdot {\bf v}_2)
+\frac{9}{2}({\bf v}_2\cdot {\bf n})^3
\bigg] \,.
\label{2PNeomfinal}
\eea
\label{eomfinal}
\ees
}
The equation of motion of body 2 can be found by interchanging $1
\rightleftharpoons 2$.
These agree
completely with other results for the 2PN equations of motion
\cite{DD81,kopeikin85,GK86,schaferwex93,bfp98,futamase01,dire2}.

The remainder of this paper provides details.  In Sec. \ref{sec:basic}
we lay out the basic equations and assumptions that underlie our
method.  Section \ref{sec:1pneom} focuses on the first post-Newtonian
approximation and verifies SEP to that order, while Sec.
\ref{sec:2pneom} extends this to second post-Newtonian order.
Concluding remarks are made in Sec. \ref{sec:conclusion}.  Selected
detailed calculations are relegated to a series of Appendices.

\section{Basic Equations and Assumptions}
\label{sec:basic}

We consider the motion of a binary system made up of two fluid balls of
characteristic mass $m$ and size $s$.  We do not assume that the bodies are
spherically symmetric.  They could be distorted, either because of rotation
or because of tidal interactions (although we will not take either spin or
tidal effects explictly into account).  The only symmetry we will impose on
the bodies is that each be symmetric on reflection through its center of
mass.  In practice this means that we will ignore any
odd-$\ell$ angular harmonics in its density distribution; equivalently we will
assume that the integral of any {\em odd} number of purely internal 
vectors (such
as position or velocity relative to the center of mass) over the body
vanishes.

We will treat the material making up the bodies 
as a perfect fluid, with the pressure required for
equilibrium provided by random internal fluid motions.  Thus we will break
up the velocity of each fluid element in the A-th body 
according to ${\bf v} = {\bf v}_A +
{\bar {\bf v}}$, where ${\bf v}_A$ will be a suitably defined center-of-mass
velocity, and ${\bar {\bf v}}$ could contain both random thermal-like
velocities as well as bulk internal or rotational velocities.  We set the
formal pressure equal to zero ($p=0$).
We also characterize the fluid by the so-called ``conserved'' baryon mass
density,
$\rho^*$, given by
\be
\rho^* =m n \sqrt{-g}u^0 \,,
\ee
where $m$ is the rest mass per baryon, $n$ is the baryon number
density, $u^\alpha$ is the four-velocity, and
$g \equiv \det (g_{\mu\nu})$ is the determinant of the spacetime metric
$g_{\mu\nu}$.  
Henceforth, we use units in which $G=c=1$; Greek indices range over
spacetime values $0,\,1,\,2,\,3$, while Latin indices range over spatial
values $1,\,2,\,3$.  From the conservation of baryon number 
(expressed covariantly as $\nabla_\alpha (nu^\alpha)=0$), 
$\rho^*$ satisfies the {\it exact} continuity equation
\be
\frac{\partial \rho^*}{\partial t} + \nabla \cdot (\rho^* {\bf v}) =0 \,,
\ee
where $v^\alpha = u^\alpha/u^0 = (1, {\bf v})$ 
is the fluid's coordinate velocity, and
spatial gradients and dot products use a
Cartesian metric.  In terms of $\rho^*$, the stress-energy tensor takes the
form
\begin{equation}
T^{\alpha\beta} = \rho^* (-g)^{-1/2} u^0 v^\alpha v^\beta \,.
\label{Trhostar}
\end{equation}
We define the baryon rest mass, center of
baryonic
mass, velocity
and acceleration of each body by the formulae
\begin{eqnarray}
m_A &\equiv & \int_A \rho^* d^3x \,,
\nonumber \\
{\bf x}_A & \equiv & (1/m_A) \int_A \rho^* {\bf x} d^3x \,,
\nonumber \\
{\bf v}_A & \equiv & d{\bf x}_A /dt = (1/m_A) \int_A \rho^* {\bf v} d^3x
\,,
\nonumber \\
{\bf a}_A & \equiv & d{\bf v}_A /dt = (1/m_A) \int_A \rho^* {\bf a} d^3x
\,,
\label{rhostardefinitions}
\end{eqnarray}
where we have used the general fact, implied by the equation of continuity
for
$\rho^*$, that
\begin{equation}
\frac{\partial}{\partial t} \int \rho^*(t, {\bf x}^\prime) f(t, {\bf
x},{\bf
x}^\prime) d^3x^\prime = \int \rho^*(t, {\bf x}^\prime) \left ( 
\frac{\partial}{\partial t} + 
{\bf v}^\prime \cdot \nabla^\prime \right ) f(t, {\bf
x},{\bf x}^\prime) d^3x^\prime \,.
\label{continuity2}
\end{equation}
The definitions of mass and center of mass  are not unique, of course (for a
review, see \cite{damour300}).  For example, we could have used an
effective density that included internal kinetic and gravitational potential
energies, as in Sec. 6.2 of \cite{tegp}, for example.  
As it turns out, our assumption that only even $\ell$ multipoles
of the internal density distribution matter guarantees that all such
alternative definitions actually coincide with our baryonic definition, by
symmetry.  Nevertheless, one could generalize our assumptions and consider
the effect of different center-of-mass definitions, but that would be beyond
the scope of this paper.  

To 2PN order, the  
equations of hydrodynamics have the form [II, Eq. (2.23), (2.24)]
\begin{equation}
dv^i /dt = U^{,i} + a_{PN}^i + a_{2PN}^i \,,
\end{equation}
where 
\begin{subequations}
\begin{eqnarray}
a_{PN}^i &=&
	v^2 U^{,i} -4 v^i v^j U^{,j} - 3v^i \dot U - 4 U U^{,i} 
	+ 8 v^j V^{[i,j]} 
	+ 4 \dot V^i + \half \ddot X^{,i} + \frac{3}{2} \Phi_1^{,i}
	-\Phi_2^{,i} \,, \label{eomfluid1PN}\\
a_{2PN}^i &=&
 4 v^i v^j v^k V^{j,k} +  v^2 v^i \dot U  
+ v^i v^j ( 4 \Phi_2^{,j} - 2 \Phi_1^{,j} - 2 \ddot X^{,j} )
- \half v^2 (2 \Phi_2^{,i} + \Phi_1^{,i} -  \ddot X^{,i} )
\nonumber \\
&&
+ v^j v^k ( 2 \Phi_1^{jk,i} - 4 \Phi_1^{ij,k} + 2 P_2^{jk,i} -4 P_2^{ij,k})
+v^i ( 3 \dot \Phi_2 - \half \dot \Phi_1 - \frac{3}{2} \stackrel{(3)}{X}
	+4 V^k U^{,k} )
\nonumber \\
&&
+ v^j ( 8 V_2^{[i,j]} 
	- 16 \Phi_2^{[i,j]} 
	+ 4 \ddot X^{[i,j]} + 32 G_7^{[i,j]}
	- 16 U V^{[i,j]} 
	- 4 \Sigma^{,[i}(v^{j]}v^2) 
	+ 8 V^i U^{,j} 
\nonumber \\
&&
	- 4 \dot \Phi_1^{ij} 
	- 4 \dot P_2^{ij})
%%%%%%%%%%%%%%%%%%%%%%%%%% no v below %%%%%%%%%%
	+ \frac{7}{8} \Sigma^{,i}(v^4) 
	+ \frac{9}{2} \Sigma^{,i}(v^2 U)
	- 4 \Sigma^{,i}(v^j V^j) 
	- \frac{3}{2} \Sigma^{,i}(\Phi_1) 
	- 6 U \Phi_1^{,i} 
	-2 \Phi_1 U^{,i} 
\nonumber \\
&&
	- 4 \Phi_1^{ij} U^{,j} 
	+ 8 V^j V^{j,i} 
	+ 4 V^i \dot U 
	+ 2 \dot \Sigma (v^i v^2)
	+ 4 U \Phi_2^{,i} 
	+ 4 \Phi_2 U^{,i}
	+ 8 U^2 U^{,i} 
	- \Sigma^{,i}(\Phi_2) 
\nonumber \\
&&
	+ \frac{3}{2} \Sigma^{,i}(U^2) 
	-2 U \ddot X^{,i} 
	- 2 \ddot X U^{,i}
	- 8 U \dot V^i 
	- \half \Sigma^{,i}(\ddot X )
	+ \frac{3}{4} \ddot X_1^{,i} 
	- \half \ddot X_2^{,i}
	+ 2 \stackrel{(3)}{X^{i}}
	+ \frac{1}{24} \stackrel{(4)}{Y^{,i}} 
\nonumber \\
&&
	+ 4 \dot V_2^i 
	- 8 \dot \Phi_2^i 
	- 6G_1^{,i} 
	- 4 G_2^{,i} 
	+ 8 G_3^{,i} 
	+ 8 G_4^{,i} 
	- 4 G_6^{,i} 
	+ 16 \dot G_7^i 
	- 4 P_2^{ij} U^{,j}
	- 4 H^{,i}
\,,\label{eomfluid2PN} 
\end{eqnarray}
\label{eomfluid}
\end{subequations}
where commas denote partial derivatives, overdots and the notation $(n)$
over functions denote partial time
derivatives $\partial/\partial t$, parentheses and square brackets
surrounding indices denote symmetrization and antisymmetrization,
respectively, and repeated spatial indices are summed.

Definitions of the potentials appear in Appendix \ref{app:potentials}; all
are defined using $\rho^*$.
For example, $U$ is the Newtonian potential, defined by
\be
U(t,{\bf x}) \equiv \int_{\cal M} \frac{\rho^*(t,{\bf x}^\prime)}
{|{\bf x}-{\bf x}^\prime | } d^3x^\prime \,.
\ee
where $\cal M$ is a constant-time hypersurface extending to the boundary of
the near-zone of the system.

Our task is then to calculate the acceleration of, say, body \#1
through 2PN order, using
\be
a^i_1 = (1/m_1) \int_1 \rho^* (dv^i/dt) d^3x \,.
\ee
At Newtonian order, the result is simple, namely
\bea
a^i_1 &=& - (1/m_1) \int_1 \int_1 \rho^* \rho^{*\prime}
\frac{(x-x^\prime)^i}{|{\bf x}-{\bf x}^\prime|^3} d^3x d^3x^\prime
\nonumber \\
&&
+ (1/m_1) \int_1 \rho^* d^3x \int_2 \rho^{*\prime} d^3x^\prime 
\left \{ \nabla_1^i \frac{1}{r} + 
({\bar x} -{\bar x}^\prime)^j \nabla_1^{ij} \frac{1}{r} 
+ \frac{1}{2} ({\bar x} -{\bar x}^\prime)^{jk} \nabla_1^{ijk}
\frac{1}{r} + \dots \right \} \,,
\label{anewt}
\eea
where we split the position and velocity of a given point inside each body
A according to
\bea
x^i &\equiv&  x_A^i + {\bar x}^i \,,
\nonumber \\
v^i &\equiv&  v_A^i + {\bar v}^i \,,
\label{split}
\eea
and expand the potential due to body \#2 in a Taylor series about the
centers of mass of the two bodies.
The use of multi-indices denotes products of vectorial objects; for example
$x^{ij} \equiv x^i x^j$, $\nabla^{ij} \equiv \nabla^{i}\nabla^{j}$, $x^M
\equiv x^{i_1} \dots x^{i_m}$, and so on. 

The first term in Eq. (\ref{anewt}) 
vanishes by symmetry (Newton's third law). The
first term in the second line
corresponds to the normal ``point-mass'' Newtonian acceleration,
$-m_2n^i/r^2$;
the second term vanishes by the definition of baryonic center of mass,
and the remaining terms are standard Newtonian
multipole coupling terms.  They depend on the size of the bodies as $s^n$
with $n \ge 2$.  In the limit
that the size of the bodies is negligible relative to their
separation, the latter terms are vanishingly small; these are the kinds of
terms that we will ignore throughout.

\section{Equations of motion at 1PN order}
\label{sec:1pneom}

At higher PN order, there is now the possibility of
correction terms proportional to inverse powers of $s$.  
To illustrate this, we evaluate
two terms from Eq. (\ref{eomfluid1PN})
explicitly:
\bea
\frac{1}{m_1} \int_1 \rho^* v^2 U^{,i} d^3x &=& -\frac{1}{m_1}
\int_1 \rho^* (v_1^2 + 2{\bf v}_1 \cdot {\bf {\bar v}} + {\bar v}^2)
d^3x
\nonumber \\
&& \times
\biggl [ \int_1 \frac{\rho^{*\prime} (x-x^\prime)^i}{|{\bf x} -
{\bf
x}^\prime |^3}d^3x^\prime
% \nonumber \\
% &&
+ \frac{m_2 x_{12}^i}{r^3} + \frac{m_2{\bar x}^j (\delta^{ij}
-
3n^in^j)}{r^3} + \dots \biggr ]
\nonumber \\
&=& \frac{2v_1^j}{m_1} {\cal H}_1^{ji} - m_2 v_1^2
\frac{x_{12}^i}{r^3}
- \frac{2{\cal T}_1}{m_1} \frac{m_2 x_{12}^i}{r^3}
+ O(s^{1/2})\,,
\label{pnterm1}
\eea
and
\bea
\frac{1}{m_1} \int_1 \rho^* UU^{,i} d^3x &=&
-\frac{1}{m_1}
\int_1 \rho^* d^3x \biggl [
\int_1 \frac{\rho^{*\prime\prime}}{|{\bf x} - {\bf x}^{\prime\prime} |}
d^3x^{\prime\prime}
+\frac{m_2}{r} - \frac{m_2}{r^3} {\bf {\bar x}} \cdot {\bf x}_{12} + \dots
\biggr ]
\nonumber \\
&&  \times \biggl [  \int_1 \frac{\rho^{*\prime} (x-x^\prime)^i}{|{\bf x}
-{\bf x}^\prime |^3}d^3x^\prime
+ \frac{m_2 x_{12}^i}{r^3} - \frac{m_2{\bar x}^j (\delta^{ij} -
3n^in^j)}{r^3} + \dots \biggr ]
\nonumber \\
&&
= - \frac{\Omega_1^{ij}}{m_1}\frac{m_2 x_{12}^j}{r^3}
+ 2\frac{\Omega_1}{m_1}\frac{m_2 x_{12}^i}{r^3}
- \frac{m_2^2 x_{12}^i}{r^4}
+O(s)\,,
\label{pnterm2}
\eea
where
${\cal T}_1$,
$\Omega_1^{ij}$,
$\Omega_1$ and
${\cal H}_1^{ij}$ are defined in Appendix \ref{app:virial}.
Note that, because ${\bar v}^2 \sim m/s$ for a body in equilibrium, the
``kinetic energy'' ${\cal T}_1$ scales as $1/s$, as do the ``gravitational
potential energy'' quantities $\Omega_1$ and $\Omega_1^{ij}$.  
The ``gravitomagnetic'' style quantity
${\cal H}_1^{ij}$ scales as ${\bar v}/s^2 \sim 1/s^{5/2}$.  We have used our
parity assumption to eliminate all terms that involve an integral over an
odd number of internal vectors on a given body.  
Also, we have not kept
terms that scale as positive powers of $s$.  One example is a term from Eq.
(\ref{pnterm1}) proportional to ${\bar x}{\bar v} \sim s^{1/2}$; for
rotating bodies it  gives spin-dependent effects, which we treated in
Papers III and IV \cite{dire3,dire4}.  Here we ignore spin.

In the combination of 1PN terms $ 4{\dot V}^i
+\frac{1}{2} {\ddot X}^{,i}$ in Eq. (\ref{eomfluid1PN}),
the time derivatives generate accelerations
inside the potentials.
To the order needed for a 1PN calculation, we must therefore
substitute the Newtonian hydrodynamic
equations for those accelerations and carry out the same procedures for the
integrals as described
above.  When we go to 2PN order, we will have to insert the 1PN hydrodynamic
equations.

Carrying out these procedures for all the terms in Eq. (\ref{eomfluid1PN}),
and dropping terms scaling as positive powers of $s$, we obtain
for the equation of motion of body 1,
\bea
a_{1PN}^{i} &=& -\frac{m_2}{r^2} n^i + \frac{m_2}{r^2} n^i \left [
4\frac{m_2}{r} + 5\frac{m_1}{r} - v_1^2 +4 {\bf v}_1 \cdot {\bf v}_2
- 2 v_2^2 + \frac{3}{2}({\bf v}_2\cdot {\bf n})^2 \right ]
\nonumber
\\
&& + \frac{m_2}{r^2} (v_1 - v_2)^i [4 ({\bf v}_1\cdot {\bf n}) -3
({\bf v}_2\cdot {\bf n})] - \frac{m_2 n
^i}{m_1 r^2} \bigg [2 {\cal T}_1 + \Omega_1 \bigg] \nonumber \\
&& + 4 \frac{m_2 n^j}{m_1 r^2} \bigg [2 {\cal T}_1^{ij} +
\Omega_1^{ij} \bigg ] + 3 \frac{n^j}{r^2} 
\bigg [2 {\cal T}_1^{ij} + \Omega_1^{ij} \bigg ] +
\frac{3}{2} \frac{n^i n^j n^k}{r^2} 
\bigg [2 {\cal T}_2^{jk} + \Omega_2^{jk} \bigg ]
\nonumber \\
&& - \frac{n^i}{r^2} \bigg[4 {\cal T}_2 + \frac{5}{2} \Omega_2
\bigg] - \frac{v_1^j}{m_1 } \bigg [ 4 {\cal H}_1^{(ij)} 
-3 {\cal K}_1^{ij} \bigg ] \,.
\label{1PNself}
\eea
Note that, apart from the ``point'' mass terms, only terms scaling as
$s^{-1}$ and $s^{-5/2}$ arise.

We now assume that each body is in equilibrium, implying that it
is stationary, or possibly periodic on an internal dynamical timescale.  As
a consequence, any time
derivative of the moment of inertia tensor,
$I^{ij} = \int\rho^* \bar x^i \bar x^j d^3 x$,
can be set to zero or can be averaged to zero.
This results in a set of
virial relations derived in Appendix \ref{app:virial}, which imply
for each body that
$2 {\cal T}_A^{ij} + \Omega_A^{ij} = 0$, 
$2 {\cal T}_A + \Omega_A= 0$, and  
$4 {\cal H}_A^{(ij)}-3 {\cal K}_A^{ij} =0$.  These 
eliminate most terms
dependent upon the structure of the bodies, leaving only the term 
$-n^i r^{-2} (4 {\cal T}_2 + \frac{5}{2} \Omega_2)$.  This cannot be eliminated by
a virial relation; however we can add an arbitrary amount of $2 {\cal T}_2 +
\Omega_2 = 0$ to it to put it into the form
\be
- \frac{n^i}{r^2} \bigg[(4-2 \alpha) {\cal T}_2 + 
(\frac{5}{2} - \alpha) \Omega_2
\bigg ] \,.
\label{body2self}
\ee
Despite these terms, we can 
make the 1PN equations independent of any $O(s^{-1})$
self terms by redefining the
masses of each of the bodies to be,
\be
M_A \equiv m_A + (4-2 \alpha) {\cal T}_A + \left (\frac{5}{2} - \alpha
\right ) \Omega_A
\label{1pnmassredef}
\ee
resulting in Eq. (\ref{1PNeomfinal}).
Note that the redefinition of the masses from $m_A$ to $M_A$ 
in the 1PN terms will affect the
equations of motion
only at 2PN order.  Thus we have verified the SEP to 1PN order.

The choice $\alpha = 3/2$ gives a redefined mass $M_A = m_A + {\cal T}_A +
\Omega_A$, which naturally represents the total (baryonic plus kinetic plus
gravitational) mass of the body, while the choice $\alpha=2$ gives $M_A =
m_A + \frac{1}{2}\Omega_A$, which is the same thing, {\it after}
applying the virial relation ${\cal T}_A = -\Omega_A/2$, and represents rest
mass plus gravitational binding energy.
Either definition, when applied to an isolated gravitating system, gives the
total system mass, as measured using Keplerian orbits far from the system.
At 1PN order, the choice of $\alpha$ is
completely arbitrary, but we will see that, to satisfy SEP at 2PN order, 
we must choose $\alpha=2$.

\section{Equations of motion at 2PN order}
\label{sec:2pneom}

We now apply the same methods at 2PN order.
We restrict
attention to terms analogous to those that arose at 1PN order, namely terms 
that scale as $s^{-1}$, $s^{-5/2}$, as well as ``point-mass'' terms.  
The most straightforward terms to evaluate are those that involve
{\em two-body} potentials and superpotentials, 
such as $U$, $V^i$, $\Phi_1$, $X$, and their various generalizations, such
as $V_2^i$, $X^i$, $Y$, etc (see Appendix \ref{app:potentials} for
definitions).   We use the splitting of position
and velocity as in Eq. (\ref{split}), combined with the scaling ${\bar v} 
\sim
s^{-1/2}$ and ${\bar x} \sim s$.  It is a simple matter to expand and sort
the terms using algebraic software.  To illustrate the
results, we cite a few simple examples from Eq. (\ref{eomfluid2PN}):
\bes
\bea
\frac{4}{m_1} \int_1 \rho^* v^iv^jv^kV^{j,k} d^3x &=& - 4 \frac{m_2}{r^2} v_1^i ({\bf v_1} \cdot {\bf n})
({\bf v_1} \cdot {\bf v_2})
- \frac{8}{m_1} \frac{m_2}{r^2} \left (v_1^i v_2^j n^k {\cal T}_1^{jk}
 + ({\bf v_1} \cdot {\bf n})v_2^j {\cal T}_1^{ij}
 +({\bf v_1} \cdot {\bf v_2})n^j {\cal T}_1^{ij} \right )
 \nonumber \\
&& + \frac{4}{m_1} (v_1^i v_1^2 {\cal K}_1 + v_1^j v_1^2 {\cal H}^{ij} )
\,,
\\
\frac{8}{m_1} \int_1 \rho^* U^2 U^{,i} d^3x &=& - 8\frac{m_2^3 n^i}{r^4} + 
16  \frac{m_2^2}{m_1 r^2} (2 n^i \Omega_1
- n^j \Omega_1^{ij} ) \,.
\eea
\ees

Throughout, we 
apply the Newtonian virial relations, $2{\cal T}_A^{ij} + \Omega_A^{ij} =0$
and $4{\cal H}_A^{(ij)} -3 {\cal K}_A^{ij} =0$ to all 2PN terms
that scale as $s^{-1}$ and $s^{-5/2}$.  In some terms, an additional
virial relation is required, involving $d^4 I_A^{ij}/dt^4$.  
An example is the term in Eq. (\ref{eomfluid2PN}) involving
$\partial^4 Y^{,i}/\partial t^4 $,
where $Y$ is the superduperpotential 
(see Appendix \ref{app:potentials} for
definitions).
Splitting $Y({\bf x})$ into two terms, one from body 1 and one from body 2,
one can show, using a Taylor expansion of the second term about the center
of mass of body 2, that, inside body 1, $Y$ is given by
\be
Y({\bf x}) = \int_1 \rho^{*\prime} |{\bf x}-{\bf x}^\prime |^3 d^3x^\prime
 + m_2 |{\bf x}-{\bf x}_2 |^3 + \frac{1}{2} I_2^{jk} \nabla^{jk} |{\bf
 x}-{\bf x}_2 |^3 + \frac{1}{12} I_2^{jklm} \nabla^{jklm} |{\bf
  x}-{\bf x}_2 |^3 + \dots \,.
\ee
Partial time derivatives of the terms involving the moments of body 2 
either will give zero, because of our virial
assumption that the moments are effectively 
constant in time, or will give terms proportional
to the moments themselves, which therefore scale as positive powers of $s$.
Thus in this case, only the contribution from body 1 and the point mass term
from body 2 will contribute; no self-terms from body 2 will arise.

Next in complexity are the so-called {\em triangle} potentials, such as
$P_2^{ij} = P(U^{,i} U^{,j})$,  and the potentials $G_a$ ($a=1..6$) and
$G_7^i$,
which depend on the field point and on {\em
two} source points, hence the name triangle potentials.  
Luckily these can all be written in analytic form using the ``triangle''
function ${\cal G}(xAB)$, which satisfies the differential equation
$\nabla^2 {\cal G}(xAB) = -(|{\bf x}-{\bf x}_A||{\bf x}-{\bf x}_B|)^{-1}$,
and which is given explicitly by
\begin{eqnarray}
{\cal G}(xAB) &\equiv& - \ln \Delta(xAB) + 1 \,,
\nonumber \\
\Delta(xAB) &\equiv& |{\bf x}-{\bf x}_A|+
	|{\bf x}-{\bf x}_B|+
		|{\bf x}_A-{\bf x}_B| \,.
		\label{GDeltadef}
\eea
For example, the potential $G_1 = P({\dot U}^2)$ can be expressed in the
form
\be
G_1 =  \sum_{A,B} \int_A \int_B \rho_A^*  \rho_B^* d^3x_A d^3x_B
v_A^iv_B^j \nabla_A^i \nabla_B^j {\cal G}(xAB) \,.
\ee
Consequently, terms involving triangle potentials can be evaluated with
ease (using algebraic software, to be sure) using the same splitting and
expansion procedure as 
before.  For example,
\bea
-\frac{6}{m_1} \int_1 \rho^*  G_1^{,i} d^3x &=& 
\frac{m_2}{r^2} \biggl \{ [ 6n^i({\bf n} \cdot {\bf v}_2)^2 - 
3n^iv_2^2 -3v_2^i ({\bf n} \cdot {\bf v}_2) ]\frac{m_2}{r} 
\nonumber \\
&&
+ [  3n^i({\bf v}_1 \cdot {\bf v}_2)
-12 n^i({\bf n} \cdot {\bf v}_1)({\bf n} \cdot {\bf v}_2) 
+6v_1^i ({\bf n} \cdot {\bf v}_2)
+3v_2^i ({\bf n} \cdot {\bf v}_1) ]
\frac{m_1}{r} \biggr \}
\nonumber \\
&& + \frac{12}{m_1}\frac{m_2}{r^2} ({\bf n} \cdot {\bf v}_2) (v_1^i \Omega_1
- v_1^j \Omega_1^{ij} ) - 6 \frac{n^i}{r^2} (v_2^2 \Omega_2 - v_2^j v_2^k
\Omega_2^{jk} ) \,.
\eea

The most difficult term to evaluate is the 
final term in Eq. (\ref{eomfluid2PN}), 
$-4H^{,i}$.  This involves the so-called ``quadrangle'' potential, since 
$H = P(U^{,ij} P_2^{ij} )$ is a function of the field point and {\em three}
source points.  Unfortunately there is no known analytic formula similar to
the function $\cal G$ that can be
employed to simplify this potential or to expand it about the bodies'
centers of mass using algebraic software.  Instead an alternative and
cumbersome method must be used; this is detailed in Appendix
\ref{app:H}.  The result for this term is
\bea
- \frac{4}{m_1} \int_1 \rho^* H^{,i} d^3x &=&
-\frac{m_2 n^i}{r^4} (8m_1m_2 +m_2^2) 
\nonumber \\
&& -\frac{m_2}{r^3} \biggl \{ (5m_1-m_2) n^i \frac{\Omega_1}{m_1} 
+(5m_1+m_2) n^j \frac{\Omega_1^{ij}}{m_1} 
-2(10m_1-m_2) n^in^jn^k \frac{\Omega_1^{jk}}{m_1} 
\nonumber \\
&&
+2 n^i \Omega_2
+2(m_1+3m_2) n^j \frac{\Omega_2^{ij}}{m_2} 
-2(m_1+6m_2) n^in^jn^k \frac{\Omega_2^{jk}}{m_2} 
\biggr \} \,.
\label{Htermresult}
\eea
In the limit of spherically symmetric bodies, where $\Omega_A^{ij} = \Omega_A
\delta^{ij}/3$, all the $s^{-1}$ contributions cancel, and
the result agrees with that of \cite{kopeikin85}.

We also must insert the 1PN hydrodynamical equations of motion into the
accelerations that appear in the 1PN terms 
$ 4{\dot V}^i
+\frac{1}{2} {\ddot X}^{,i}$ in Eq. (\ref{eomfluid1PN}), and evaluate those
2PN terms.  They involve only 2-body potentials, and thus are straightforward
to evaluate.

Combining all the terms that scale as $s^{-1}$ and $s^{-5/2}$, 
and displaying only those terms,
we obtain the result 
\bea
a_{1(2PN)}^i &=&
%begin mb n^i/ma/r^2
\frac{m_2}{r^2} \bigg[
n^i \biggl \{ 
\left ( v_1^2 + \frac{5}{2} \frac{m_1}{r} + 5\frac{m_2}{r} \right ) 
\frac{\Omega_1}{m_1}
+ \left ( v_2^2 - \frac{1}{2} v_1^2 +2{\bf v}_1 \cdot {\bf v}_2 
+ \frac{3}{4} ({\bf n} \cdot {\bf v}_2 )^2 
+ 4 \frac{m_2}{r} + 13\frac{m_1}{r} \right )
\frac{\Omega_2}{m_2}
\nonumber \\
&&
+ \left ( 4v_2^k - \frac{1}{2} v_1^k \right ) v_1^j \frac{\Omega_1^{jk}}{m_1}
+ \left ( \frac{13}{4}v_2^j v_2^k  - \frac{3}{2}v_2^2 n^jn^k
   -\frac{15}{2} \frac{m_1}{r} n^jn^k \right ) 
   \frac{\Omega_2^{jk}}{m_2}
- \frac{9}{4} v_2^j v_2^k n^l n^m  \frac{\Omega_2^{jklm}}{m_2}
\biggr \}
\nonumber \\
&&
- n^j \biggl \{ \left ( 4  {\bf v}_1 \cdot {\bf v}_2 + 
20 \frac{m_2}{r} \right ) \frac{\Omega_1^{ij}}{m_1}
+ \left ( 3v_2^2 +15 \frac{m_1}{r} \right ) \frac{\Omega_2^{ij}}{m_2}
\biggr \}
\nonumber \\
&&
+ (v_1 -v_2)^i \biggl \{ \left ( 2{\bf n} \cdot {\bf v}_1
- \frac{3}{2}{\bf n} \cdot {\bf v}_2 \right ) \frac{\Omega_2}{m_2}
+ 4 v_1^j n^k \frac{\Omega_1^{jk}}{m_1} \biggr \}
\nonumber \\
&&
+ \left ( 4 ({\bf n} \cdot {\bf v}_1)v_2^j - ({\bf n} \cdot {\bf v}_1)v_1^j
-({\bf n} \cdot {\bf v}_2)v_1^j \right )
\frac{\Omega_1^{ij}}{m_1}
\nonumber \\
&&
- 9 v_1^j v_1^k n^l \frac{\Omega_1^{ijkl}}{m_1}
-\frac{9}{2} v_2^j v_2^k n^l \frac{\Omega_2^{ijkl}}{m_2}
-\frac{1}{m_1} v_1^j v_1^k v_1^l 
\left ( \frac{15}{2} {\cal K}_1^{ijkl} -3 {\cal L}_1^{ijkl} -6 {\cal
L}_1^{jikl} \right )
\bigg ] \,,
\label{2PNeomself}
\eea
where $\Omega_A^{ijkl}$, which scales as $s^{-1}$, and ${\cal
K}_1^{ijkl}$ and ${\cal L}_1^{ijkl}$, which scale as $s^{-5/2}$, are
defined in Appendix \ref{app:virial}.

We now must return to the 1PN equations of motion including $s^{-1}$ and
$s^{-5/2}$ terms, Eq. (\ref{1PNself}), and employ virial
relations that are correct to 1PN order.
As we saw in Sec \ref{sec:1pneom}, all self terms vanish at 1PN
order, except for the term proportional to $4{\cal T}_2 + 5/2 \, \Omega_2$.
The residual term there could be absorbed into a redefinition of the masses.
We must now consider the application of virial relations
and the mass redefinition
at higher PN order.  

We first replace the term
$-(n^i/r^2)(4{\cal T}_2 + 5/2 \, \Omega_2)$ in Eq.  (\ref{1PNself})
by the equivalent term
\be
- \frac{n^i}{r^2} \bigg[(4-2 \alpha) {\cal T}_2 + 
(\frac{5}{2} - \alpha) \Omega_2
\bigg ]  - \alpha \frac{n^i}{r^2} (2{\cal T}_2 + \Omega_2 )\,,
\label{selfpiece}
\ee
and apply the 1PN corrected virial relations of 
Eqs. (\ref{1PNvirial}) only to the second piece of this
expression, as well as to all the other $s^{-1}$ and
$s^{-5/2}$ terms in Eq. (\ref{1PNself}).  
The other piece of (\ref{selfpiece})
will be absorbed into a redefinition of the
mass of body 2.
At the same time, we redefine all the masses in the point-mass 1PN terms
using Eqs. (\ref{1pnmassredef}); for those terms, to the order of
approximation needed, the Newtonian virial
relations may be used to simplify the renormalization to $M_A = m_A + 1/2 \,
\Omega_A$.   The $s^{-1}$ and
$s^{-5/2}$ terms generated by these substitutions cancel all the $s^{-5/2}$
terms and {\em almost} all the
$s^{-1}$ terms in Eq. (\ref{2PNeomself}).  The uncancelled $s^{-1}$ terms at
1PN and 2PN order, combined with the Newtonian acceleration, give
\bea
a_{1}^i({\rm self}) &=&
%begin mb n^i/ma/r^2
-\frac{n^i}{r^2} \biggl [
m_2 + (4-2\alpha){\cal T}_2 + \left ( \frac{5}{2} -\alpha \right ) \Omega_2
\nonumber
\\
&& + (\alpha -2)v_2^2 \Omega_2 - \frac{1}{2} (21 - 10 \alpha)
\frac{m_1}{r}\Omega_2 - \frac{1}{4} (13 - 6 \alpha) v_2^j v_2^k
\Omega_2^{jk} \biggr ]
\,.
\eea
Note that the choice $\alpha =2$ leaves the coefficient of the ``Newtonian''
acceleration as
\bea
m_2 + \frac{1}{2} \Omega_2 - \frac{1}{2}\frac{m_1}{r}\Omega_2 
- \frac{1}{4} v_2^j v_2^k \Omega_2^{jk} 
&=& m_2 + \frac{1}{2} {\hat \Omega}_2
\nonumber \\
&=&
M_2 \,,
\label{2pnmass}
\eea
where 
\be
{\hat \Omega}_2 = \left ( 1 - \frac{m_1}{r} \right )\Omega_2
- \frac{1}{2} v_2^j v_2^k \Omega_2^{jk}
\label{omegahat}
\ee
is the gravitational binding energy 
{\em as calculated in the local, comoving inertial frame} of body 2
(see Appendix \ref{app:mass} for derivation).
The quantity $M_2$ is precisely the total mass, comprising the
baryonic mass $m_2$ plus the locally measured 
gravitational binding energy $\frac{1}{2} {\hat \Omega}_2$.  The 
additional terms in (\ref{omegahat})
are simply the $s^{-1}$ corrections arising from the 
transformation from the local, comoving inertial frame to our global
coordinate frame.    

Thus, when all masses are written in terms of the new 
$M_A$, no $s^{-1}$ terms
survive in the equations of motion, leaving only the ``point''-mass terms,
given by  Eq. (\ref{2PNeomfinal}).

\section{Conclusions}
\label{sec:conclusion}

We have studied the motion of finite sized, self-gravitationally
bound, arbitarily shaped bodies at second post-Newtonian order, and
demonstrated that self-energy effects that scale with size of the body
as $s^{-1}$ and $s^{-5/2}$ cancel precisely when a suitable
renormalization of the masses is carried out.  Since the bodies are
finite, that renormalization is finite (in contrast to other, singular
renormalization techniques), and consists of redefining the masses, to
the PN order needed, as the sum of baryonic rest mass and
gravitational binding energy, {\em as measured in the locally comoving
inertial frame of each body}.  We emphasize that we have made no
effort to define masses or centers of mass in a covariant manner; all
calculations are carried out in the harmonic coordinates that are
built into our approach.  We are currently investigating 2PN terms with
other scalings, such as $s^{-7/2}$, $s^{-2}$ and so on, to see if they also
cancel.

\acknowledgments
This work is supported in part by the National Science Foundation,
Grant Nos. PHY 03-53180 and PHY 06-52448.  
One of us (CMW) is grateful to the Group
Gravitation Relativiste et Cosmologie (GR$\varepsilon$CO) of the Institut
d'Astrophysique de Paris for its hospitality while this work was being
completed.

%#########
\appendix

\section{Potentials appearing in the 2PN equations of motion}
\label{app:potentials}

The potentials that appear in the equations of motion
are all Poisson-like potentials and their generalizations, namely a
superpotential $X$ and a superduperpotential $Y$,
integrated over a constant time hypersurface
$\cal M$ that extends to the boundary of the near zone of the system.
In the case of integrands of non-compact support, we
discard all terms that depend on the radius of the
near-zone, $\cal R$; see Paper I \cite{dire1} for justification.
All potentials are defined in terms of the conserved
baryon mass density $\rho^*$:
\begin{eqnarray}
P(f) &\equiv& \frac{1}{4\pi} \int_{\cal M} 
\frac{f(t,{\bf x}^\prime)}{|{\bf x}-{\bf x}^\prime | }
d^3x^\prime \,, \quad \nabla^2
P(f) = -f \,, \nonumber \\
\Sigma (f) &\equiv& \int_{\cal M} \frac{\rho^*(t,{\bf x}^\prime)f(t,{\bf
x}^\prime)}{|{\bf x}-{\bf x}^\prime | } d^3x^\prime = P(4\pi\rho^* f) \,,
\nonumber \\
X(f)  &\equiv& \int_{\cal M} {\rho^*(t,{\bf x}^\prime)f(t,{\bf
x}^\prime)}
{|{\bf x}-{\bf x}^\prime | } d^3x^\prime \,,
\nonumber \\
Y(f) &\equiv& \int_{\cal M} {\rho^*(t,{\bf x}^\prime)f(t,{\bf
x}^\prime)}
{|{\bf x}-{\bf x}^\prime |^3 } d^3x^\prime \,.
\label{definesuper}
\end{eqnarray}

The specific potentials used in the 1PN and 2PN equations of
motion are then given by
\begin{eqnarray}
U &\equiv& \Sigma(1) \,,  \qquad V^i \equiv \Sigma(v^i) \,, \qquad  
\Phi_1^{ij} \equiv \Sigma(v^iv^j)
\,,
\nonumber \\
\Phi_1 &\equiv& \Sigma(v^2) \,, \qquad
\Phi_2 \equiv \Sigma(U) \,, \qquad  X \equiv X(1) \,,
\nonumber \\
V_2^i &\equiv& \Sigma(v^iU) \,, \qquad  \Phi_2^i \equiv \Sigma(V^i) \,,
\qquad Y \equiv Y(1) \,,
\nonumber \\
X^i &\equiv& X(v^i) \,, \qquad  
X_1 \equiv  X(v^2) \,, \qquad  X_2 \equiv X(U) \,,
\nonumber \\
P_2^{ij} &\equiv& P(U^{,i}U^{,j}) \,, \qquad  P_2 \equiv
P_2^{ii}=\Phi_2
-\frac{1}{2}U^2 
\,,\nonumber \\
G_1 &\equiv& P({\dot U}^2)  \,, \qquad  G_2 \equiv P(U {\ddot U}) \,,
\nonumber \\
G_3 &\equiv& -P({\dot U}^{,k} V^k) \,, \qquad  G_4 \equiv
P(V^{i,j}V^{j,i}) \,,\nonumber \\
G_5 &\equiv& -P({\dot V}^k U^{,k}) \,, \qquad  G_6 \equiv P(U^{,ij}
\Phi_1^{ij}) \,,\nonumber \\
G_7^i &\equiv& P(U^{,k}V^{k,i}) + \frac{3}{4} P(U^{,i}\dot U ) \,,
\qquad  H \equiv P(U^{,ij} P_2^{ij}) \,.
\label{potentiallist}
\end{eqnarray}
We refer the reader to Ref. \cite{dire2}, Appendix C, for further discussion
of the triangle and quadrangle potentials.

\section{Virial Theorems}
\label{app:virial}

\subsection{Newtonian virial relations}
\label{app:Nvirial}

We assume that our bodies are in equilibrium, so that they
are either stationary, or at worst periodic on an internal dynamical timescale.
This implies, among other things, that any time 
derivative of the moment of inertia tensor, 
$I^{ij} = \int\rho^* \bar x^i \bar x^j d^3 x$, 
is either zero or averages to zero. This will give us several virial
relations 
that will simplify our equations.  Considering body 1, for example, we
have that
\bes
\bea
\frac{1}{2} 
\dot I_1^{ij} &=& \int_1\rho^* \bar v^{(i} \bar x^{j)} d^3x  \label{idot}
\,, \\
\frac{1}{2} 
\ddot I_1^{ij} &=&\int_1\rho^* (\bar v^i \bar v^j + \bar x^{(i} a^{j)}  
)d^3 x \label{iddot} 
\,, \\
\frac{1}{2} 
\dddot I_1^{ij} &=&\int_1\rho^* (3\bar v^{(i} a^{j)} + \bar x^{(i} \dot{a}^{j)} 
 ) d^3 x \,,
\label{idddot} 
\eea
\label{idots}
\ees
where, by virtue of the fact that $\int_1 \rho^* \bar x^i d^3x =\int_1 \rho^*
\bar v^i d^3x = 0$, we can drop the bars on the accelerations.
Substituting the Newtonian equation of motion for $a^i$ gives
\bes
\bea
\frac{1}{2} 
\ddot I_1^{ij} &=&
 \int_1\rho^* \bar v^i \bar v^j d^3 \bar x 
-\int_1\int_1 \rho^*\rho^{* \prime}
\frac{ \bar x^{(i} ( x- x^\prime)^{j)} }
{| {\bf x} -  {\bf x}^\prime|^3} d^3 x d^3 x^\prime 
+\int_1 \rho^*
\bar x^{(i} U_2^{,j)} d^3x
\,,
\nonumber \\
&=& 2 {\cal T}_1^{ij} +  \Omega_1^{ij}  + O(s^2)
\label{ddotI}
\,, \\
\frac{1}{2} 
\dddot I_1^{ij} &=& 
- \int_1\int_1\rho^*\rho^{* \prime} 
\bigg (
3 \frac{\bar v^{(i} ( x -  x^\prime)^{j)}}
{| {\bf x} -  {\bf x}^\prime|^3} 
+\frac{\bar x^{(i} ( v -  v^\prime)^{j)}}
{| {\bf x} -  {\bf x}^\prime|^3} 
- 3 \frac{ \bar x^{(i} ( x -  x^\prime)^{j)}
( {\bf x} -  {\bf x}^\prime) \cdot ( {\bf v} -  {\bf v}^\prime)}
{| {\bf x} -  {\bf x}^\prime|^5} \bigg ) d^3 x d^3 x^\prime  \nonumber \\
\nonumber \\
&& + \int_1 \rho^* \biggl ( 3\bar v^{(i} U_2^{,j)} + \bar x^{(i}
\frac{d}{dt} U_2^{,j)} \biggr ) d^3x 
\nonumber \\
&=& 4 {\cal H}_1^{(ij)} - 3 {\cal K}_1^{ij} + O(s^{1/2}) \,,
\label{dddotI}
\eea
\label{dotsI}
\ees
where
\bea
{\cal T}_1^{ij} & \equiv& \frac{1}{2}\int_1\rho^* \bar v^i \bar v^j d^3x 
\,,
\nonumber \\
{\cal T}_1 & \equiv& {\cal T}_1^{ii} = \frac{1}{2}\int_1\rho^* \bar v^2 d^3x 
\,,
\nonumber \\
\Omega_1^{ij} & \equiv& -\frac{1}{2}\int_1\int_1 \rho^*\rho^{* \prime}
\frac{(  x- x^\prime)^{ij}}
{|{\bf x} - {\bf x}^\prime|^3} d^3 x d^3 x^\prime 
\,,
\nonumber \\
\Omega_1 & \equiv& \Omega_1^{ii} = -\frac{1}{2}\int_1\int_1 \frac{\rho^*\rho^{* \prime}}
{|{\bf x} - {\bf x}^\prime|} d^3 x d^3 x^\prime 
\,,
\nonumber \\
{\cal H}_1^{ij} & \equiv& \int_1\int_1 \rho^*\rho^{* \prime}
\frac{{v^\prime}^i ( x - x^\prime)^j}{| {\bf x} -  {\bf x}^\prime|^3} 
d^3 x d^3 x^\prime 
\,,
\nonumber \\
{\cal K}_1^{ij} & \equiv& \int_1\int_1 \rho^*\rho^{* \prime}
\frac{ ( x - x^\prime)^{ij} {\bf v}^\prime \cdot 
( {\bf x} - {\bf x}^\prime)}{| {\bf x} -  {\bf x}^\prime|^5} 
d^3 x d^3 x^\prime \,,
\eea
and where we expanded the potential $U_2$ in the same manner as in Eq.
(\ref{anewt}), yielding only terms that scale as positive powers
of $s$.  (In Paper III \cite{dire3}, 
the $s^{1/2}$ terms in Eq. (\ref{dddotI}) produced
some spin-orbit terms that contributed to the 1PN spin-orbit equations
of motion.)  Recall that, in these definitions, $( x - x^\prime)^{ij\dots}
\equiv ( x - x^\prime)^i ( x - x^\prime)^j \dots$.

Setting the derivatives of $I^{ij}$ to zero, we obtain the $s^{-1}$
and $s^{-5/2}$ virial relations for body 1,
\bea
2 {\cal T}_1^{ij} + \Omega_1^{ij} &=&  0 \,,
\nonumber \\
4 {\cal H}_1^{(ij)} - 3 {\cal K}_1^{ij} &=& 0 \,.
\label{virialNewt}
\eea

\subsection{Post-Newtonian virial relations}
\label{app:PNvirial}

Working at 2PN order in the equations of motion requires us to obtain our
virial relations correct to
1PN order.  Accordingly we must now substitute the 1PN
hydrodynamical equations into Eqs. (\ref{idots}).  
The resulting 1PN corrections will have terms with a variety of
scalings, from purely internal terms scaling as $s^{-2}$ in $\ddot
I^{ij}$ and $s^{-7/2}$ in $\dddot I^{ij}$, to terms arising from
expansion of the potentials due to body 2, also with a variety of scalings.
Here we focus only on contributions scaling as $s^{-1}$ and
$s^{-5/2}$.  A straightforward calculation then yields the 1PN virial
relations,
\bes
\bea
0 &=& 2 {\cal T}_1^{ij} + \Omega_1^{ij} - v_1^2 \Omega_1^{ij}
- \frac{3}{2} v_1^k v_1^l \Omega_1^{ijkl} 
-5 \frac{m_2}{r} \Omega_1^{ij} \,,
\\
0&=&
4 {\cal H}_1^{(ij)} - 3 {\cal K}_1^{ij}
+ \frac{15}{2} v_1^k v_1^l{\cal K}_1^{ijkl}
- 9 v_1^k v_1^l {\cal L}_1^{(ijk)l} 
\nonumber \\
&& 
- \frac{m_2}{r^2} ({\bf v} \cdot {\bf n}) \Omega_1^{ij}
+ 4 \frac{m_2}{r^2} (v_1-2v_2)^k n^{(i} \Omega_1^{j)k}
- 8 \frac{m_2}{r^2} v^{(i} \Omega_1^{j)k}n^{k} 
+3 \frac{m_2}{r^2} v_1^k n^l \Omega_1^{ijkl} \,,
\eea
\label{1PNvirial}
\ees
where ${\bf v} \equiv {\bf v}_1 - {\bf v}_2$, and 
\bea
\Omega_1^{ijkl} & \equiv& -\frac{1}{2}\int_1\int_1 \rho^*\rho^{* \prime}
\frac{(  x- x^\prime)^{ijkl}}
{|{\bf x} - {\bf x}^\prime|^5} d^3 x d^3 x^\prime
\,,
\nonumber \\
{\cal K}_1^{ijkl} & \equiv& \int_1\int_1 \rho^*\rho^{* \prime}
\frac{ ( x - x^\prime)^{ijkl} {\bf v}^\prime \cdot
( {\bf x} - {\bf x}^\prime)}{| {\bf x} -  {\bf x}^\prime|^7}
d^3 x d^3 x^\prime \,,
\nonumber \\
{\cal L}_1^{ijkl} & \equiv& \int_1\int_1 \rho^*\rho^{* \prime}
\frac{ {v^\prime}^i ( x - x^\prime)^{jkl}}{| {\bf x} -  {\bf x}^\prime|^5}
d^3 x d^3 x^\prime \,.
\eea
The virial relations for body 2 can be obtained from these by
the interchange $1 \rightleftharpoons 2$, with 
${\bf n} \to -{\bf n}$. 

\section{Renormalized mass in body's rest frame}
\label{app:mass}

We have renormalized the masses of the bodies by defining the total mass to
be a sum of baryonic mass and gravitational binding energy, namely
$M_A \equiv m_A + \frac{1}{2} \Omega_A$ modulo 2PN corrections.  However,
those 2PN corrections are of order $m\epsilon^2 \sim m(m/s)^2$, and thus scale
as $s^{-2}$.  As we are only looking at $s^{-1}$ terms, we have not kept
those corrections.  However, there is an additional, 2PN order, $s^{-1}$
correction to the total mass that must be taken into account.

The baryonic mass of each body is a scalar, frame-invariant quantity, but
the gravitational binding energy is not, since it depends on the size of the
body.  The correct, frame-invariant definition of our total mass must
therefore be that mass as measured in a local inertial frame momentarily
comoving with the body.  Thus we define
\be
M_A \equiv m_A + \frac{1}{2} {\hat \Omega}_A + \delta M_A \,,
\ee
where $\delta M_A$ denotes the 2PN, $s^{-2}$ corrections, which we are
ignoring, and 
\be
{\hat \Omega}_A \equiv  -\frac{1}{2}
\int_A\int_A \frac{{\hat \rho}^* {\hat \rho}^{*\prime}}
{|{\hat {\bf x}}- {\hat {\bf x}}^\prime|} d^3{\hat x} d^3{\hat x}^\prime 
\,,
\label{omegahatdef}
\ee
where hats denote spatial variables defined on a constant time hypersurface
in the comoving frame.   
Now, the quantity ${\hat \rho}^* d^3{\hat x}$ is invariant, but 
$|{\hat {\bf x}}- {\hat {\bf x}}^\prime|$ is not.  In Paper II, Appendix B,
we showed that the transformation between spatial coordinates ${\hat x}^i$
in the comoving
frame of body A
and our global harmonic coordinates $x^i$ takes the form
\be
x^i = x_A^i + {\hat x}^j \left \{ \delta_j^i + \epsilon (A^i_j 
+B^i_{jk} {\hat x}^k )\right \} + O(\epsilon^2) \,,
\label{localcoord}
\ee
where $A^i_j$ and $B^i_{jk}$ are functions of the 
basis transformation from the global frame to the comoving inertial
frame,
\be
{\vec e}_\mu = (\Lambda_\mu^{\hat \alpha} + {\tilde B}_{\mu\nu}^{\hat
\alpha} {\bar x}^\nu ) {\vec e}_{\hat \alpha} \,,
\ee
where ${\bar x}^\nu = x^\nu - x_A^\nu$. The coefficients
$\Lambda_\mu^{\hat \alpha}$ correspond to boosts and coordinate
rescalings, and the ${\tilde B}_{\mu\nu}^{\hat
\alpha}$ make the frame freely falling.

Substiting Eq. (\ref{localcoord}) into the definition of $\Omega_1$,
we obtain
\be
\Omega_1 = {\hat \Omega}_1 - \epsilon A^{ij} {\hat \Omega}_1^{ij}
+ O(\epsilon^2) {\hat \Omega}_1 \,,
\ee
where the term arising from $B^i_{jk}$ produces integrals over body 1
involving an odd number of vectors, which therefore vanish.
The coefficients $A_{ij}$ arise from a boost to the velocity $v_1^i$
of body 1, combined with a rescaling to an orthonormal basis in the
field of body 2.  To the first PN order needed, it is simple to show
that 
\be
A^{ij} = -\frac{1}{2}v_1^i v_1^j - \frac{m_2}{r}\delta^{ij}  \,.
\ee
The result is, to the required order,
\be
\Omega_1 = {\hat \Omega}_1 + \frac{1}{2} v_1^{i}v_1^{j} {\hat \Omega}_1^{ij}
+ \frac{m_2}{r} {\hat \Omega}_1 \,.
\ee
Therefore, to 2PN order, and with $s^{-1}$ scaling, the {\it total}
mass of each body is given by
\bea
M_A &=& m_A + \frac{1}{2} {\hat \Omega}_A + O(s^{-2}) 
\nonumber \\
&=& m_A + \frac{1}{2} \Omega_A - \frac{1}{4}  v_A^{i}v_A^{j} \Omega_A^{ij}
-\frac{1}{2} \frac{m_B}{r} \Omega_A  + O(s^{-2})\,, 
\eea
where we drop the hats on the 2PN terms.

\section{The quadrangle potential $H$}
\label{app:H}

We now turn to
evaluation of the term involving the quadrangle potential $H = P(U^{,ij}
P_2^{ij})$.  This potential can be written in the form
\bea
H &=& \frac{1}{4\pi} \int_{\cal M}
	\frac{d^3x^\prime}{|{\bf x}-{\bf x}^\prime|}
		U^{,ij}(x^\prime) P_2^{ij}(x^\prime)
		\nonumber \\
&=& \sum_{ABC}
\int_A \rho_A^* \nabla_A^i \nabla_A^j d^3x_A
\int_B \rho_B^* \nabla_B^i d^3x_B
\int_C \rho_C^* \nabla_C^j d^3x_C
\, {\cal H}(xA;BC) \,,
\label{Hdef}
\eea
where the function ${\cal H}$ of four field points is defined by
\begin{equation}
{\cal H}(AB;CD) \equiv \frac{1}{(4\pi)^2} \int_{\cal M}\int_{\cal M}
\frac{d^3x^\prime d^3x^{\prime\prime}}
{|{\bf x}_A-{\bf x}^\prime|
|{\bf x}_B-{\bf x}^\prime|
|{\bf x}^\prime-{\bf x}^{\prime\prime}|
|{\bf x}_C-{\bf x}^{\prime\prime}|
|{\bf x}_D-{\bf x}^{\prime\prime}|} \,.
\label{calHdef}
\end{equation}
Unfortunately, there appears to be no closed-form analytic expression for
${\cal H}$ similar to Eq. (\ref{GDeltadef}) for $\cal G$.  Instead, we use
the first definition of $H$ in Eq. (\ref{Hdef}), and integrate $H^{,i}$ over
the mass of body 1.  After integrating once by parts, showing that the
surface term at the boundary of $\cal M$ can be discarded, and using the
fact that $P_{2 \,\,,j}^{jk}= \frac{1}{2} \Phi_2^{,k} - \frac{1}{2} UU^{,k}
-\Sigma (U^{,k})$, we obtain
\be
\int_1 \rho^* H^{,i} d^3x = \frac{1}{4\pi} \int_{\cal M} U_1^{,ij} U^{,k} P_2^{jk}
d^3x + \frac{1}{8\pi} \int_{\cal M} U_1^{,i} U^{,k} [\Phi_2^{,k}-UU^{,k}
-2\Sigma (U^{,k}) ] d^3x \,,
\label{gradH}
\ee
where $U_1$ is the Newtonian potential due to body 1 only.  The first term
in Eq. (\ref{gradH}) can be expanded into the form
\be
{\rm Term}_1 = \frac{1}{4\pi} \int_{\cal M}  U_1^{,ij} ( U_1^{,k}+ U_2^{,k})
[ P_{2(11)}^{jk}+2P_{2(12)}^{(jk)}+P_{2(22)}^{jk} ] d^3x \,,
\label{Hterm1}
\ee
where the subscripts denote the contributions from the various bodies.

The second term in Eq. (\ref{gradH}) can be simplified by integrating by
parts, leading to 
\be
{\rm Term}_2 = \frac{1}{4} \int_1 \rho^* (U^{,i}\Phi_2 + U\Phi_2^{,i} -
U^2U^{,i} )d^3x + \frac{1}{4} \int_{\cal M} \rho^* U_1^{,i} \Phi_2 d^3x
-  \frac{1}{4\pi} \int_{\cal M} U_1^{,i} U^{,k} \Sigma (U^{,k}) d^3x \,,
\label{Hterm2}
\ee
where the first integral is only over body 1.

To handle the integrals of noncompact support integrands, we split the
domain $\cal M$ into three regions, a region ${\cal M}_1$ 
of radius ${\cal R}_1$
surrounding body 1 and a region ${\cal M}_2$ 
of radius ${\cal R}_2$ surrounding body 2,
and the remainder, ${\cal M} - {\cal M}_1 -{\cal M}_2$.
To carry out the integrals, we will need suitable forms for the various
potentials, $U$, $P_{2(11)}^{jk}$,  $P_{2(12)}^{(jk)}$, and so on, in the
appropriate regions.  

This will be aided by a general expansion of
the function ${\cal G}(ABC)$ in powers of
$r_{AB}/r_{AC}$, where points $A$ and $B$ are assumed to lie inside
one body, and point $C$ is inside the other body, so that $r_{AB} \sim s
\ll r_{AC}$.
Straightforward
methods lead to the expansion
\begin{eqnarray}
{\cal G}(ABC) &=& -\ln r_{AC} + 1 - \ln 2
+ \frac{1}{2} \sum_{m=0}^\infty
\frac{(- r_{AB})^{m+1}}{(m+1)!}
\nonumber \\
&&
\times  \biggl \{ ({n}_{AB})^M \nabla_A^M \left ( \frac{1}{r_{AC}}
\right)
+ \frac{r_{AC}}{m+1} ({n}_{AB})^{M+1}
\nabla_A^{M+1} \left ( \frac{1}{r_{AC}} \right )
\biggr \}
\,.
\label{calGsum}
\end{eqnarray}
where $n_{AB}^i = x_{AB}^i/r_{AB}$.
Then, given that 
\be
P_{2(AB)}^{ij} =  \int_A \int_B \rho_A^*  \rho_B^*
          d^3x_A d^3x_B \nabla_A^i \nabla_B^j {\cal G}(xAB) \,,
\ee
it can be shown that, for two source points in body A and a field
point outside the body,
\be
P_{2(AA)}^{ij} = \frac{1}{4} \frac{m_A^2}{y_A^2} ({\hat y}_A^i {\hat
y}_A^j -\delta^{ij})
+ \frac{1}{y_A} (\Omega_A^{ij} - \Omega_A\delta^{ij}) 
+ O(s) \,,
\label{P2AA}
\ee
where ${\hat y}_A^i \equiv (x-x_A)^i/|{\bf x}-{\bf x}_A|$,
and $y_A \equiv |{\bf x}-{\bf x}_A|$, and $x_A^i$ now denotes the
center of baryonic mass of body A.  For a spherically symmetric body,
this agrees with Eqs. (C6) and (C7) of Paper II \cite{dire2}. 

Similarly, for a source point and a field point in body A, and the
other source point in body B, 
\bea
P_{2(AB)}^{ij} &=& \frac{m_Am_B}{2r^2} (\delta^{ij}-2n^{ij})+
\frac{m_B}{2r^2} n^j X_A^{,i}
\nonumber \\
&&
+\frac{3}{4} \frac{m_B}{r^3} n^{<jk>} \left ( 2X_A \delta^{ik} -
\frac{1}{3}Y_A^{,ik} \right )
+ \frac{1}{4} \frac{m_Am_B}{r^3} {\bar x}^k (4n^{ijk} 
-n^i\delta^{jk}
-n^k \delta^{ij} -2n^j \delta^{ik} )
\nonumber \\
&&
+\frac{1}{12} \frac{m_B}{r^4} n^{<jkl>} \left ( Z_A^{,ikl} -
15Y_A^{,k} \delta^{il} \right )
\nonumber \\
&&
-\frac{1}{4} \frac{m_B}{r^4} \left ( m_A {\bar x}^{kl} + I_A^{kl} \right )
\left ( 6n^{ijkl} -2n^{ik}\delta^{jl}-2n^{kl}\delta^{ij}
-4n^{jk}\delta^{il} + \delta^{ik}\delta^{jl} \right ) 
\nonumber \\
&&
+ O(s^3) \,,
\label{P2AB}
\eea
where $n^i=(x_A^i-x_B^i)/r$, $r=|{\bf x}_A-{\bf x}_B|$, 
and ${\bar x}^k = x^k - x_A^k$;
$X_A$, $Y_A$, and $Z_A$ denote the superpotentials $\int_A
\rho^{*\prime} |{\bf x}-{\bf x}^\prime |^p d^3x^\prime$ generated by
body A only, where $p=1,\, 3 ,\, 5$, respectively, and $I_A^{kl}$ is
the moment of inertia tensor of body A.  
Angular brackets $<>$ around indices denote the symmetric trace-free
product.
We have kept terms up to
order $s^2$ in Eq. (\ref{P2AB}) because they will ultimately be
multiplied by terms that scale as negative powers of $s$.

For a source point in body A, a source point in body B
and the field point between the two bodies,
\be
P_{2(AB)}^{ij} =  \frac{m_Am_B}{\Delta(xAB)}
\left ( \frac{({\hat y}_A-n)^i ({\hat y}_B+n)^j}{\Delta(xAB)}
+ \frac{\delta{ij}-n^{ij}}{r} \right ) + O(s) \,.
\ee
Other useful identities involving integrals over a sphere 
surrounding one of the bodies, say body B, include,
\bes
\bea
\frac{1}{4\pi} \int_0^{{\cal R}_B} \frac{d^3x}{|{\bf x}-{\bf x}_A||{\bf x}-{\bf
x}_B|} &=& 
- \frac{1}{3{\cal R}_B} {\bf x}_A \cdot {\bf x}_B
- \frac{1}{2} r_{AB}+ {\cal R}_B  \,,
\label{identity1}
\\
\frac{1}{4\pi} \int_0^{{\cal R}_B} 
\frac{x^i d^3x}{|{\bf x}-{\bf x}_A||{\bf x}-{\bf x}_B|} 
&=& - \frac{1}{15{\cal R}_B} \left [x_B^2 x_A^i + x_A^2 x_B^i -
3(x_A^i+x_B^i){\bf x}_A \cdot {\bf x}_B \right ]
- \frac{1}{4}(x_A^i+x_B^i)r_{AB} 
\nonumber \\
&&+ \frac{{\cal R}_B}{3} (x_A^i+x_B^i) \,.
\label{identity2}
\eea
\label{identities}
\ees

To illustrate the method used in evaluating
$\int_1 \rho^* H^{,i}d^3x$, we consider one integral in Eq.
(\ref{Hterm1}),
namely
$(4\pi)^{-1} \int_{\cal M} U_1^{,ij} U_1^{,k} P_{2(22)}^{jk} d^3x$.
Considering first the integral over a sphere of radius ${\cal R}_1$
surrounding body 1, we expand $P_{2(22)}^{jk}$ in powers of ${\bar
x}^m = x^m - x_1^m$
about the center of mass of body 1.  The only term that gives a
non-zero result that scales as $s^0$ or lower is the term linear in
$\bar x$.  We must therefore evaluate the integral
$\int  U_1^{,ij} U_1^{,k} {\bar x}^m d^3x$ over a sphere surrounding
body 1.  This can be done using Eqs. (\ref{identities}).  Combining the result
with ${P_{2(22),m}^{jk}}$ evaluated at $x_1$ using Eq. (\ref{P2AA}), and
keeping terms scaling as $s^{-1}$ or ${\cal R}_1^{-1}$,
we obtain the term
\be
\frac{m_2^2}{4r^3} ( \Omega_1 n^i - \Omega_1^{ij} n^j + 2
\Omega_1^{jk} n^{ijk} ) + \frac{m_1^2}{30 {\cal R}_1 r^2}
\left ( \frac{7m_2^2}{r}n^i + 14 \Omega_2 n^i - 2\Omega_2^{ij} n^j  \right ) 
\,.
\label{term1}
\ee
Integrating over a sphere surrounding body 2, we expand the product 
$ U_1^{,ij} U_1^{,k}$ about $x_2$ in powers of ${\bar x}^m = x^m-x_2^m$.
But because, inside body 2, $P_{2(22)}^{jk}$ scales as $s^{-2}$, the
integral $\int_2 {\bar x}^M P_{2(22)}^{jk} d^3x$ scales as $s^{m+1}$,
yielding no $s^{-1}$ or $s^0$ terms of interest.  Finally, to integrate over
the domain ${\cal M}-{\cal M}_1-{\cal M}_2$, we note that, because the
integral over the domain ${\cal M}_2$ yields only positive power scaling,
then for our purposes, we can integrate over this domain equally well
by evaluating the integral over the
exterior of the sphere surrounding body 1,
$\int_{{\cal R}_1}^\infty U_1^{,ij} U_1^{,k} P_{2(22)}^{jk} d^3x$, with the
external potential $U_1 = m_1/r$, and the exterior form (\ref{P2AA}) for 
$ P_{2(AA)}^{jk}$, but using body 2 as the source.  The fact that the
chosen form of
$ P_{2(22)}^{jk}$ is singular at $x_2$ does not affect the parts of the
integral we are interested in, and the integral is finite.  
The result for this term is
\be
-\frac{7}{30{\cal R}_1} \frac{m_1^2 m_2^2}{r^3}n^i
-\frac{1}{15{\cal R}_1} \frac{m_1^2}{r^2} (7\Omega_2 n^i - \Omega_2^{ij} n^j
) 
+ \frac{1}{4} \frac{m_1^2}{r^3} (\Omega_2 n^i - \Omega_2^{ij} n^j
+ 2 \Omega_2^{jk} n^{ijk} ) \,.
\label{term2}
\ee
Combining expressions (\ref{term1}) and (\ref{term2}), we see that the terms
proportional to ${\cal R}_1^{-1}$ cancel, as they must, leaving the result for
this term,
\be
\frac{1}{4} \frac{m_1^2}{r^3} (\Omega_2 n^i - \Omega_2^{ij} n^j
+ 2 \Omega_2^{jk} n^{ijk} ) 
+ \frac{1}{4} \frac{m_2^2}{r^3} (\Omega_1 n^i - \Omega_1^{ij} n^j
+ 2 \Omega_1^{jk} n^{ijk} ) \,.
\ee
This happens to be antisymmetric on interchange of 1 with 2 (whereby $n^i
\to - n^i$), which is to be expected, since this particular term can also be
written in the form
\be
\frac{1}{2(4\pi)^2} \int_{\cal M} U_1^{,j} U_1^{,k} {U_2^{,j}}^\prime 
{U_2^{,k}}^\prime \frac{d^3x d^3x^\prime (x-x^\prime)^i}
{|{\bf x}-{\bf x}^\prime |^3} \,,
\ee
which is manifestly antisymmetric under $1 \rightleftharpoons 2$.
The remaining contributions to Term$_1$ in Eq. (\ref{Hterm1}) and
Term$_2$ in Eq. (\ref{Hterm2}) can be evaluated in the same manner.
The final result for terms scaling as $s^{-1}$, together with the
point-mass contributions, is given by Eq. (\ref{Htermresult}).

\end{document}